\documentclass[10pt,aps,prb,twocolumn,longbibliography,superscriptaddress,floatfix]{revtex4-2}

\usepackage{bm}
\usepackage{upgreek}
\usepackage{amsmath}
\usepackage{graphicx}
\usepackage{nicefrac}
\usepackage{siunitx}
\usepackage[capitalize]{cleveref}

\begin{document}

\title{Phonon angular momentum transfer torque}

\author{Verena Brehm} 
\affiliation{Department of Applied Physics and Science Education,
Eindhoven University of Technology, 5612 AP Eindhoven, Netherlands}
\author{Daniel A. Bustamante Lopez} 
\affiliation{Department of Applied Physics and Science Education,
Eindhoven University of Technology, 5612 AP Eindhoven, Netherlands}
\affiliation{Department of Physics, University of Fribourg, CH-1700 Fribourg, Switzerland}
\author{Shu Zhang}
\affiliation{Okinawa Institute of Science and Technology Graduate University, Onna-son, Okinawa 904-0412, Japan}
\author{Dominik Juraschek}
\email{dominik.juraschek@unifr.ch}
\affiliation{Department of Applied Physics and Science Education,
Eindhoven University of Technology, 5612 AP Eindhoven, Netherlands}
\affiliation{Department of Physics, University of Fribourg, CH-1700 Fribourg, Switzerland}
\affiliation{Paul Scherrer Institut, CH-5232 Villigen PSI, Switzerland}

\begin{abstract}
 
Angular momentum in solids is carried by (quasi)particles such as magnons, plasmons, and phonons, and coupling between these reservoirs enables diverse hybrid phenomena. Here, we introduce the transfer of angular momentum from lattice vibrations to magnetic order through the phonon angular momentum transfer torque (PAMTT), generated by a polarized phonon bath. We identify thermal routes for generating phonon angular momentum: a phonon Edelstein accumulation in noncentrosymmetric crystals and temperature-gradient-induced PAM currents with longitudinal Seebeck-like and transverse Hall components. Using realistic order-of-magnitude estimates, we find that PAMTT could produce measurable ferromagnetic-resonance frequency shifts and, for favorable interfacial coupling, drive magnetization reversal on nanosecond timescales.

\end{abstract}

\maketitle

\section*{Introduction}

Controlling transport and transfer of angular momentum lies at the heart of spintronic applications. Well-established electrical means of controlling magnetization include spin-transfer, spin-orbit, and orbital torques, in which a spin- or orbital-polarized electric current transfers angular momentum to an adjacent magnetic layer \cite{Manchon2019,Fert2024,Fukami2025}. As an alternative to electrical control, it is desirable to create and transport pure spin currents in insulating materials without electrical currents, as studied in the field of spin insulatronics \cite{arneSpinInsulatronics}. In recent years, lattice vibrations (phonons) have also emerged as alternative carriers of angular momentum, with vibrational atomic displacements following elliptical trajectories \cite{Juraschek2025}. 
Here, we formulate a phonon angular momentum transfer torque (PAMTT) for thermally generated, incoherent phonon angular momentum, as a lattice analogue of spin-transfer torque. This provides a route to manipulate magnetization in insulating heterostructures that combine magnetic components with noncentrosymmetric, nonmagnetic chiral-phonon materials and enable interconversion between spin and lattice angular momentum.

\section*{Theoretical formalism}\label{sec:LLG-PAMTT}

We adapt the Landau-Lifshitz-Gilbert-Slonczewski equation \cite{theOriginalSlonczewski} to include a PAMTT at the interface between a bulk crystal and a thin-film magnetic insulator,
\begin{align}\label{eq:LLG-PAMTT}
        \partial_t \mathbf{m}_i = &\frac{\gamma_{\mathrm{el}}}{1+\kappa_{\mathrm{el}}^2} \left[ \mathbf{m}_i\times \mathbf{B}^0_i + \kappa_{\mathrm{el}} \mathbf{m}_i\times \left(\mathbf{m}_i \times \mathbf{B}^0_i\right) \right] \\ &- \tau L \left[\beta_{\mathrm{P}} \mathbf{m}_i \times \mathbf{l}^{\mathrm{ph}} + \beta_{\mathrm{R}} \mathbf{m}_i \times \left(\mathbf{m}_i \times \mathbf{l}^{\mathrm{ph}}\right)\right],  \nonumber
\end{align} 
where $\mathbf{m}_i$ is the normalized magnetic moment, $\gamma_{\mathrm{el}}<0$ is the electron gyromagnetic ratio, $\kappa_{\mathrm{el}}$ denotes the Gilbert damping, and $\mu_s$ is the atomistic magnetic moment. The local PAM accumulation entering the torque is $\mathbf{L}^{\mathrm{ph}}=L\mathbf{l}^{\mathrm{ph}}$, where $L=|\mathbf{L}^{\mathrm{ph}}|$ is its magnitude and $\mathbf{l}^{\mathrm{ph}}$ its unit polarization direction.
In the first line of \cref{eq:LLG-PAMTT}, $\mathbf{B}^0_i = - \frac{1}{\mu_s}\frac{\partial \mathcal{H}^0}{\partial \mathbf{m}_i}$ denotes the effective field derived from a generic spin Hamiltonian, with $\mathcal{H}^0 = \mathcal{H}_{\mathrm{ex}} + \mathcal{H}_{\mathrm{aniso}} + ...$ representing the magnetic order. 

The second line of \cref{eq:LLG-PAMTT} describes the PAMTT. Microscopically, the PAM of an atom of mass $m_{\mathrm{ion}}$ and displacement $\mathbf{u}$ is $\mathbf{L}^{\mathrm{ph}}=m_{\mathrm{ion}}\mathbf{u}\times\dot{\mathbf{u}}$ \cite{Juraschek2025}. 

Furthermore, we introduce dimensionless scaling factors for the precession component, $\beta_{\mathrm{P}}$, and the relaxation component, $\beta_{\mathrm{R}}$. 
For spin-transfer and spin-orbit torques, scaling factors for the relaxation and precession components are highly interface- and material-dependent and are determined by the microscopic origin of the coupling \cite{Manchon2019}. We therefore treat $\beta_{\mathrm{P}}$ and $\beta_{\mathrm{R}}$ as free parameters. %


\section*{Generation and transfer of PAM}\label{sec:originPAMaccumulation}

\begin{figure*}
    \centering\includegraphics[width=0.98\linewidth]{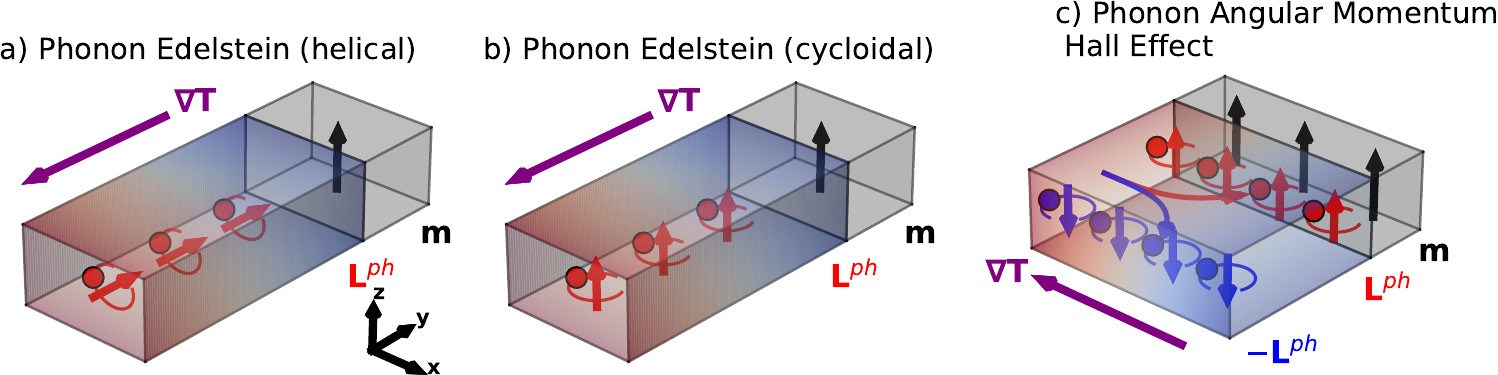}
    \caption{Geometries for PAMTT. A temperature gradient, indicated by the red-to-blue color gradient, generates a PAM accumulation or current according to (a, b) the phonon Edelstein effect \cite{HamadaJfromTempGrad,Zhang2025} and (c) the phonon angular momentum Hall effect \cite{ParkPAMHE,PAMHEwithDaniel}. At the interface, the local PAM $\mathbf{L}^{\mathrm{ph}}=L\mathbf{l}^{\mathrm{ph}}$ exerts a torque on the magnetization $\mathbf{m}$ in the adjacent magnet. For the temperature-gradient directions shown here, the PAM is either perpendicular or parallel to the magnetization.}
    \label{fig:setups}
\end{figure*}

We consider two fundamental mechanisms for generating a local PAM accumulation \cite{HamadaJfromTempGrad,Zhang2025,ParkPAMHE,PAMHEwithDaniel}, schematically shown in \cref{fig:setups}. First, in the phonon Edelstein effect, a heat current in a noncentrosymmetric material drives the phonon population out of equilibrium and creates an imbalance between modes carrying opposite PAM, producing a bulk accumulation. Second, in the phonon angular momentum Hall effect, a temperature gradient generates a PAM current that can build up an accumulation where it varies spatially or terminates, for example at a boundary. These two responses can be written as
\begin{equation}\label{eq:JalphaT}
    \frac{L_i^{\mathrm{E}}}{V_c}=\alpha_{ik}\nabla_k T,\qquad J^{\mathrm{ph}}_{i,j}=\sigma_{ijk}\nabla_k T .
\end{equation}
Here $L_i^{\mathrm{E}}$ is the Edelstein PAM accumulated within a unit cell of volume $V_c$, so $L_i^{\mathrm{E}}/V_c$ is the corresponding PAM density, while $J^{\mathrm{ph}}_{i,j}$ is the current density of PAM component $i$ flowing along direction $j$. The tensors $\bm{\upalpha}$ and $\bm{\upsigma}$ describe the PAM-accumulation response and PAM-current conductivity, respectively. The first relation in \cref{eq:JalphaT} describes the phonon Edelstein effect. In the second, $j=k$ describes longitudinal, Seebeck-like PAM transport, whereas $j\neq k$ describes transverse, Hall-like transport. The local $\mathbf{L}^{\mathrm{ph}}$ entering \cref{eq:LLG-PAMTT} may therefore contain both direct Edelstein and current-induced contributions.

\subsection*{Phonon Edelstein effect}\label{subsec:TempGrad}

Hamada et al.\ \cite{HamadaJfromTempGrad} estimated the response tensor $\bm{\upalpha}$ for representative polar and chiral crystals. The relevant tensor components are of comparable order of magnitude, but the tensor form, and therefore the direction of the induced PAM, depends on crystal symmetry. In the polar case, the induced PAM is orthogonal to the temperature gradient, whereas in the helical case it can be parallel. Together with the magnetic anisotropy, this determines whether the PAM can be arranged parallel or perpendicular to the magnetization, as illustrated in \cref{fig:setups}a,b.

Writing $\lambda^{\mathrm{ph}}$ for the phonon relaxation time, Ref.~\cite{HamadaJfromTempGrad} reports $|\alpha_{yx}|/\lambda^{\mathrm{ph}}$ for wurtzite GaN and $|\alpha_{zz}|/\lambda^{\mathrm{ph}}$ and $|\alpha_{xx}|/\lambda^{\mathrm{ph}}$ for helical Te to be on the order of $10^{-7}\,\si{\joule\per\meter\squared\per\kelvin}$ at $T=\SI{300}{\kelvin}$, see Appendix~\ref{A:alphasFromHamada} for details. Comparable values were reported for 2D halide perovskites \cite{MikeChiralPhin2DPerovskites}. Thus, we estimate $L^{\mathrm{E}}$ to be on the order of $10^{-4}\hbar$, assuming $\lambda^{\mathrm{ph}}=\SI{10}{\pico\second}$, a temperature gradient of $\SI{10}{\kelvin\per\micro\meter}$, and a unit-cell volume $V_c=\SI{1}{\nano\meter\cubed}$.

\subsection*{Phonon angular momentum Hall effect}\label{sec:PAMHE}

As a second route to a local PAM accumulation, we consider the \textit{phonon angular momentum Hall effect} \cite{ParkPAMHE,PAMHEwithDaniel}, depicted in \cref{fig:setups}c. A longitudinal temperature gradient generates a transverse PAM current through mixing of different Cartesian vibrational polarizations by the force constants. This transverse response is described by the components with $j\neq k$ of $\sigma_{ijk}$ in \cref{eq:JalphaT}. Termination of this current produces the edge accumulation $L^{\mathrm{H}}$. In steady state, $L_i^{\mathrm{H}}=-V_c\lambda^{\mathrm{ph}}\partial_jJ^{\mathrm{ph}}_{i,j}$. The PAM Hall effect can occur even in highly symmetric lattices such as square or honeycomb structures \cite{PAMHEwithDaniel}. Its conversion efficiency depends on the relevant force constants: diagonal next-nearest-neighbor force constants in square lattices and nearest-neighbor force constants in honeycomb lattices. For representative systems, finite-sample calculations yield PAM edge accumulations on the order of $10^{-3}\hbar$ for graphene and $10^{-2}\hbar$ for silicon. Explicit analytical expressions can be found in the Supplementary Information of Ref.~\cite{PAMHEwithDaniel}.

\subsection*{Interfacial transfer and coupling strength}\label{subsec:quantities}

The estimates above indicate that local PAM values on the order of $L\sim10^{-4}$--$10^{-2}\hbar$ can be generated through a temperature gradient and set the angular-momentum scale available for transfer at an interface. To estimate how much of this PAM can enter the magnetic material, we refer to Suzuki et al.\ \cite{suzuki2026theoryphononangularmomentum}, who report the diffusion of PAM across interfaces between chiral and achiral crystals. Depending on the phonon dispersion and sound velocity, PAM can reduce or even increase near the interface. In all cases, PAM can penetrate into the first few layers of the adjacent material. We therefore assume that an interface can be constructed such that the estimated interfacial PAM acts locally on the magnetization through PAMTT without significant losses across the interface. 
The remaining question concerns the coupling coefficient $\tau$ between PAM and magnetization. Writing $\tau=-\mu_0\gamma_{\mathrm{ph}}\gamma_{\mathrm{el}}/V_c$, recent studies, summarized in Ref.~\cite{geilhufeReviewMuPhonon}, predict phonon magnetic moments on the order of $\mu_B$, motivating values of $|\gamma_{\mathrm{ph}}|$ that can approach $|\gamma_{\mathrm{el}}|$ in favorable cases. The PAMTT rate scale is set by $|\tau|L$, while its orientation is determined by $\mathbf{l}^{\mathrm{ph}}$ and the torque coefficients in \cref{eq:LLG-PAMTT}.

\begin{figure}
    \centering
    \includegraphics[width=\linewidth]{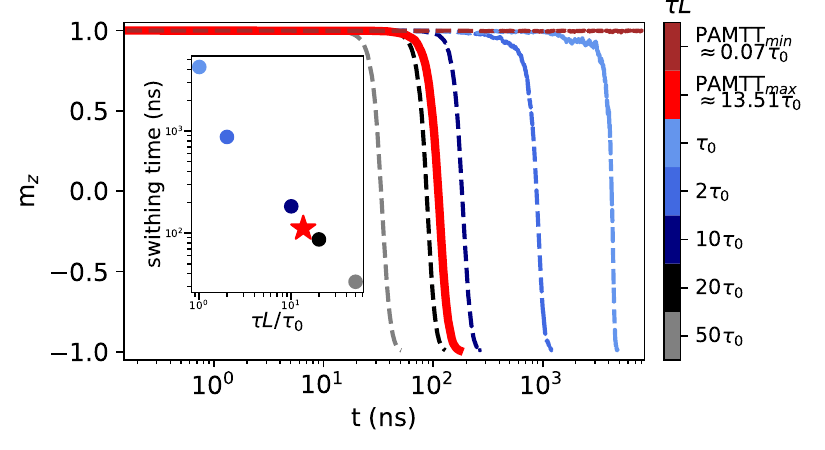}
    \caption{Switching of magnetization with PAMTT in a macrospin model. The main plot shows the magnetization component $m_z$. Switching times are summarized in the inset. For $-\mathbf{l}^{\mathrm{ph}}\parallel\mathbf{m}$ and $\beta_{\mathrm{R}}=1$, \cref{eq:LLG-PAMTT} gives the threshold for compensating the Gilbert damping, $\tau_0 = -\frac{2\kappa_{\mathrm{el}}d_z\gamma_{\mathrm{el}}}{\mu_s(1+\kappa_{\mathrm{el}}^2)}>0$, represented by the light blue line. Increasing the PAMTT strength above this scale reduces the switching time. For an optimistic estimate of PAMTT$_{\mathrm{max}}$ with $L = 10^{-2}\hbar$ and $\gamma_{\mathrm{ph}}=|\gamma_{\mathrm{el}}|$, highlighted in red, we find a switching time on the order of $10^2$~ns.}
    \label{fig:switching}
\end{figure}

\section*{Examples of induced magnetization dynamics}\label{sec:MagDynamics}

\subsection*{Modification of ferromagnetic resonance (FMR) and damping}

From \cref{eq:LLG-PAMTT}, for $\mathbf{l}^{\mathrm{ph}}\parallel\mathbf{m}$ and $\kappa_{\mathrm{el}}=0$, the precession term shifts the FMR angular frequency to
$\omega=-2d_z\gamma_{\mathrm{el}}/\mu_s+\beta_{\mathrm{P}}\tau L$, see Appendix~\ref{sec:shiftOfFMR} for a short derivation. Reversing the PAM direction reverses this frequency shift. The relaxation term similarly modifies the Gilbert damping, enhancing it for one PAM orientation and reducing it for the opposite orientation. This damping control can be understood as a modification of the decay rate of the magnetic system's magnon density and could be used to drive magnon transport in nonlocal geometries. 
Given the precision of FMR measurements, the frequency shift and the change in Gilbert damping due to PAMTT should appear as a shift and as a broadening or narrowing of the FMR peak, respectively. Since the PAM direction can be reversed by changing the temperature-gradient direction, taking the difference between $\pm\mathbf{l}^{\mathrm{ph}}$ would allow background signals to be filtered. Reversing $\mathbf{l}^{\mathrm{ph}}$ at fixed magnetization gives $\Delta\omega=2\beta_{\mathrm{P}}\tau L$, hence $\Delta\omega/(2\pi)=\beta_{\mathrm{P}}\tau L/\pi$. For example, for $L=(0.005\text{--}0.01)\hbar$, $\beta_{\mathrm{P}}=1$, $V_c=\SI{1}{\nano\meter\cubed}$, and $\gamma_{\mathrm{ph}}=(0.01\text{--}1)|\gamma_{\mathrm{el}}|$, the frequency difference is around $\SI{0.07}{\mega\hertz}$--$\SI{13}{\mega\hertz}$.

\subsection*{Magnetization switching}

For the PAM orientation that reduces the Gilbert damping, the magnetic state becomes linearly unstable when $\beta_{\mathrm{R}}\tau L>\tau_0$, where
$\tau_0\equiv-\frac{2\kappa_{\mathrm{el}}d_z\gamma_{\mathrm{el}}}{\mu_s(1+\kappa_{\mathrm{el}}^2)}$.
We estimate the resulting switching timescales by simulating macrospin dynamics via numerical integration of \cref{eq:LLG-PAMTT} using a Heun procedure. 
Phonon dynamics is not included explicitly in the simulation, but enters through $\tau L$, $\mathbf{l}^{\mathrm{ph}}$, $\beta_{\mathrm{P}}$, and $\beta_{\mathrm{R}}$. The macrospin model is described by an easy-axis anisotropy $\mathcal{H}^0 = -d_z (\mathbf{m}\cdot\hat{\mathbf{e}}_z)^2$ with $d_z=\SI{5}{\micro\electronvolt}$, $\kappa_{\mathrm{el}}=10^{-4}$, and $\mu_s=\mu_\text{B}$, representing a ferromagnet with an FMR frequency of about \SI{5}{\giga\hertz}. We set $\beta_{\mathrm{P}}=\beta_{\mathrm{R}}=1$ and include weak thermal noise corresponding to $k_B T = 0.01 d_z$, modeled as uncorrelated Gaussian white noise.
At the threshold $\tau L=\tau_0$, the finite thermal noise used here leads to switching after nearly \SI{4}{\micro\second}, as shown in \cref{fig:switching}. Increasing the PAMTT strength to $\tau L=2\tau_0$ reduces the switching time to \SI{1}{\micro\second}, while $\tau L=10\tau_0$, $20\tau_0$, and $50\tau_0$ yield switching times of \SI{180}{\nano\second}, \SI{80}{\nano\second}, and \SI{30}{\nano\second}. 
For a conservative estimate of PAMTT$_{\mathrm{min}}$ with $L = 0.005\hbar$ and $\gamma_{\mathrm{ph}}=0.01|\gamma_{\mathrm{el}}|$, indicated by the brown line, no switching occurs since the corresponding PAMTT strength is weaker than $\tau_0$. For an optimistic estimate of PAMTT$_{\mathrm{max}}$ with $L = 0.01\hbar$ and $\gamma_{\mathrm{ph}}=|\gamma_{\mathrm{el}}|$, highlighted in red, we find a switching time of around \SI{110}{\nano\second}.

\section*{Conclusion}\label{sec:Conclusion}
We propose that thermally generated phonon angular momentum accumulations and current-induced interfacial accumulations can exert a transfer torque on an adjacent magnetic layer. We suggest that PAMTT could be induced in state-of-the-art setups for detecting PAM \cite{CIPAMS,ChiralPhononActivatedSEEeffect,Zhang2025}, producing magnetization dynamics detectable with FMR and, for sufficient coupling, potentially leading to magnetization reversal. Although this study focuses on ferromagnetic systems, the same formalism can be extended to antiferromagnetic systems. By analogy with spin-transfer torque, the same coupling should also permit a reciprocal pumping process from the magnetic system to the phononic system. 
Similar to its electronic counterpart, we anticipate that PAMTT will become relevant for novel information-processing applications, including neuromorphic spintronics.


\begin{acknowledgments}
We thank Benedetta Flebus for useful discussions. This research was supported in part by grant no. NSF PHY-2309135 to the Kavli Institute for Theoretical Physics (KITP). D.M.J. acknowledges support from the ERC Starting Grant CHIRALPHONONICS, no. 101166037.
\end{acknowledgments}

\bibliography{pnas-sample}
\vspace{1cm}
\appendix
\section{Estimations of the response tensor $\alpha$} \label{A:alphasFromHamada}
For GaN (wurtzite, $P6_3mc$), Ref.~\cite{HamadaJfromTempGrad} reports $\alpha_{xy}=-\alpha_{yx}\approx 10^{-7}\times[\lambda^{ph}/1\,\mathrm{s}]\,\si{\joule\second\per\meter\squared\per\kelvin}$ at $T=\SI{300}{K}$. For Te and Se (helical structures, $P3_{121}$ or $P3_{221}$), the nonzero elements are $\alpha_{zz}\approx -10^{-7}\times[\lambda^{ph}_\parallel/1\,\mathrm{s}]\si{\joule\second\per\meter\squared\per\kelvin}$ and $\alpha_{xx}\approx -10^{-7}\times[\lambda^{ph}_\perp/1\si{s}]\si{\joule\second\per\meter\squared\per\kelvin}$ for Te, and $\alpha_{zz}\approx -10^{-6}\times[\lambda^{ph}_\parallel/1\si{s}]\si{\joule\second\per\meter\squared\per\kelvin}$, $\alpha_{xx}\approx -10^{-7}\times[\lambda^{ph}_\perp/1\si{s}]\si{\joule\second\per\meter\squared\per\kelvin}$ for Se.

\section{Shift of FMR due to PAMTT \label{sec:shiftOfFMR}} 
Here we provide a short derivation of the shift of FMR frequency under the influence of PAMTT. For a macrospin model with easy axis along the $z$-direciton, described by $\mathcal{H}^0 = -d_z (\mathbf{m}\cdot\hat{e}_z)^2$, the effective field entering the LLG is $\mathbf{B}^0 = - \frac{\partial \mathcal{H}^0}{\partial \mathbf{m}} = 2d_z \mathbf{m}\cdot\hat{e}_z$. Let's furthermore assume $\mathbf{l}^{ph}=\pm \hat{e}_z$ and neglect the Gilbert damping $\alpha=0$ and relaxation-like torque contribution $\beta_R=0$. With that, the LLG in this macrospin model simplifies to
\begin{align}\label{eq:LLG-PAMTT}
        \partial_t \mathbf{m} = -2d_z\frac{\gamma_{el}}{\mu_s}  \mathbf{m}\times \hat{e}_z \pm \tau L \beta_{P} \mathbf{m} \times \hat{e}_z .
\end{align} 
Writing out the cross products and performing linear spin wave theory assuming a magnetization along the $z$-direction $\mathbf{m} \approx (\delta_x, \delta_y,1) $, \cref{eq:LLG-PAMTT} simplifies to
\begin{align*}
    \partial_t \delta_x &= \left(-2d_z\frac{\gamma_{el}}{\mu_s}  \pm \tau L \beta_{P}\right) \delta_y \\
    \partial_t \delta_y &= -\left(-2d_z\frac{\gamma_{el}}{\mu_s}  \pm \tau L \beta_{P}\right) \delta_x
\end{align*}
which can trivially be solved by a plane wave ansatz, resulting in $\omega = 2d_z\frac{\gamma_{el}}{\mu_s} \mp \tau L \beta_{P} $.

\end{document}